\documentclass[aps,prc,groupedaddress,showpacs,twocolumn,floatfix]{revtex4}

\usepackage{graphicx}

\def\gtorder{\mathrel{\raise.3ex\hbox{$>$}\mkern-14mu
 \lower0.6ex\hbox{$\sim$}}}
\def\ltorder{\mathrel{\raise.3ex\hbox{$<$}\mkern-14mu
 \lower0.6ex\hbox{$\sim$}}}
\def\mugegm{\mu_p G_E / G_M}
\def\gegm{G_E / G_M}
\def\ge{G_E}
\def\gm{G_M}
\def\etal{\textit{et al.}}

\begin{document}

\title{Implications of the discrepancy between proton form factor measurements}

\author{J. Arrington}

\affiliation{Argonne National Laboratory, Argonne, Illinois 60439, USA}

\date{\today}

\begin{abstract}

Recent polarization transfer measurements of the proton electromagnetic form
factors yield very different results from previous Rosenbluth extractions.
This inconsistency implies uncertainties in our knowledge of the form factors
and raises questions about how to best combine data from these two techniques.
If the discrepancy is due to missing corrections to the cross section data, as
has been suggested, then the true form factors, related to the proton
structure, differ from the form factors that parametrize the deviation from
point scattering, and different applications will require the use of different
form factors. We present two extractions of the form factors: a global fit to
the world's cross section data, and a combined extraction from polarization
transfer and cross section data. The former provides a parametrization of the
elastic electron-proton cross section. The latter provides a consistent
extraction of the underlying form factors, under the assumption that missing
terms in the radiative correction explain the difference between the cross
section and polarization transfer results.

\end{abstract}
\pacs{25.30.Bf, 13.40.Gp, 14.20.Dh}

\maketitle

%%%%%%%%%%%%%%%%%%%%%%%%%%%%%%%%%%%%%%%%%%%%%%%%%%%%%%%%%%%%%%%%%%%%%%%%%%%%%

\section{INTRODUCTION}\label{sec:intro}

The proton electromagnetic form factors, $\ge$ and $\gm$, parametrize
deviations from a point particle in elastic electron-proton scattering, and are
related to the charge and magnetization distribution of the proton.  The form
factors depend only on $Q^2$, the square of the four-momentum transfer, and
until recently it was believed that the electric and magnetic form factors
showed approximate scaling, i.e. nearly identical $Q^2$
dependence~\cite{walker94}. More recent Jefferson Lab
measurements~\cite{jones00, gayou01, gayou02} utilized the polarization
transfer technique to measure the ratio $\gegm$ and found that $\ge$ decreases
more rapidly than $\gm$ at large $Q^2$.  The polarization transfer
measurements are more precise at high $Q^2$, and significantly less sensitive
to systematic uncertainties than the Rosenbluth separation measurements.
However, the two techniques disagree significantly even in the region where
both yield precise results.

At the present time, it is not known why the techniques give different
results. The systematic uncertainties of the polarization transfer measurements, primarily
spin transport and backgrounds, have been carefully studied~\cite{punjabi03}. A
detailed global analysis of the cross section measurements~\cite{arrington03}
does not show any inconsistencies in the cross section datasets, or yield any
likely candidate to explain the discrepancy.  To resolve the discrepancy,
a systematic error in the cross section would have to have a significant
dependence on the virtual photon polarization $\epsilon$, $\epsilon^{-1} = 1 +
2 (1+Q^2/4M_p^2) \tan^2{(\theta_e/2)}$, where $M_p$ is the proton mass and
$\theta_e$ is the electron scattering angle. Such a systematic error 
would have to yield a (5--7)\% $\epsilon$ dependence in the cross
section, roughly linear in $\epsilon$, in order to resolve the discrepancy.

There appear to be two possibilities: either a fundamental flaw in the
Rosenbluth or polarization transfer formalism, or an error in either the cross
section or polarization transfer measurements.  Recent works have suggested
that additional radiative correction terms, related to two-photon exchange
corrections, may lead to an error in determining the form factors from the
measured cross sections~\cite{blunden03, guichon03, rekalo03}.  If the
two-photon exchange mechanism, or some other correction that is neglected in
the cross section extraction, is the source of the discrepancy, then the form
factors extracted from a Rosenbluth separation of cross section data will
\textit{not} represent the underlying structure of the proton, but they
\textit{will} parametrize the elastic electron-proton cross section in the
usual one-photon approximation.  Conversely, the true form factors will not
yield the correct cross sections, and will thus give incorrect results if
used as a parameterization of the elastic cross section in data analysis.

If the two-photon exchange term explains the discrepancy, then the
polarization transfer result will relate to the true form factors, assuming
that the two-photon exchange has a much smaller effect on the polarization
transfer than on the Rosenbluth extractions.  However, the existing
polarization transfer experiments~\cite{jones00, gayou01, gayou02}
have extracted the ratio $\gegm$, rather than the individual form factors.  To
extract the form factors, these data must be combined with cross section
measurements to determine the absolute magnitudes of $\ge$ and $\gm$.  If the
two-photon exchange correction modifies the cross sections from those
calculated from the underlying form factors, then it is not possible to
consistently combine the two kinds of measurements without some assumption
about the two-photon exchange correction.

In this paper, we present two extractions of the proton form factors. From a
global analysis of cross section measurements, we extract the ``Rosenbluth
form factors''.  From a combined analysis of cross section and polarization
transfer data, with a ``minimal'' assumption about the nature of the
two-photon exchange corrections, we extract the ``Polarization form factors''.
If two photon corrections are the source of the discrepancy, then the
Rosenbluth form factors will parametrize the elastic cross section, and are
therefore useful as input to analysis or simulations that require the
electron-proton cross section.  The Polarization form factors will provide the
true form factors, which relate to the underlying structure of the proton.
These form factors are often described as the Fourier transformations of the
charge and magnetization distributions of the proton in the Breit frame,
although relativistic effects and the fact that each value of $Q^2$
corresponds to a different Breit frame lead to substantial theoretical
difficulties in extracting charge and magnetization
distributions~\cite{kelly02}.

\section{GLOBAL ROSENBLUTH ANALYSIS}\label{sec:xsec}

The ``Rosenbluth form factors'' are determined from a global fit to
elastic electron-proton cross section measurements. The details of the fitting
procedure are described in Ref.~\cite{arrington03}. For the present analysis
we include more recent Jefferson Lab measurements of elastic
scattering~\cite{christy03, dutta03, niculescuphd}, as well additional
datasets to constrain the low $Q^2$ behavior~\cite{borkowski74, murphy74,
simon80, simon81} to the datasets used in Ref.~\cite{arrington03,
note_resdata}. In addition, we include all of the high $Q^2$ data, up to
30~GeV$^2$, while the previous analysis was limited to 8~GeV$^2$.  The older
data have updated radiative corrections, and the small-angle data from Walker
\etal~\cite{walker94} are excluded, as described in~\cite{arrington03}.  The
form factors are fit to the following form:
\begin{equation}
\ge(Q^2), \gm(Q^2)/\mu_p = [ 1 + p_2 Q^2 + p_4 Q^4 + ... + p_{2N} Q^{2N}]^{-1},
\label{eq:fitform}
\end{equation}
where $\mu_p$ is the magnetic dipole moment of the proton and $Q^2$ values are
in GeV$^2$.  Reasonable fits are achieved for $N \geq 3$.  Note that this is a
different functional form than used in previous fits~\cite{bosted94,
arrington03, brash02}, which used polynomials in $q=\sqrt{Q^2}$.  The
polynomial in $q$ is a very general form, with adequate flexibility to
reproduce the data, but does not have the proper behavior as $Q^2 \rightarrow
0$.

The fit is quite insensitive to the order of the polynomial above $N=6$,
except for $\ge$ at large $Q^2$.  For $Q^2$ above 6~GeV$^2$, fits with nearly
identical $\chi^2$ values can have $\gegm$ either rise or fall
dramatically with $Q^2$. This is a result of the reduced sensitivity to $\ge$ 
and the limited $\epsilon$ coverage for $Q^2$ values
above 6~GeV$^2$. To avoid unreasonable behavior in the region where $\ge$ is
unconstrained by data, we keep the \textit{ratio} $\gegm$ fixed for all $Q^2$
values above 6~GeV$^2$.  This leads to a fit for $\ge$ which is continuous, but
not smooth, at $Q^2 = 6$~GeV$^2$. Because $\ge$ has relatively little
contribution to the total cross section at these momentum transfers, the cross
section extracted is still quite smooth, and the value of $\ge$ at large $Q^2$
values has little effect on the cross section, as long as the fit is
constrained to avoid $|\mu_p\ge| \gg |\gm|$.

The normalization factor for each data set is allowed to vary along with the
parameters of the fitting functions for $\ge$ and $\gm$.  The total $\chi^2$
from the cross section measurements and normalization factors is:
\begin{equation}
\chi^2_
\sigma = \sum_{i=1}^{N_\sigma} \frac{(\sigma_i - \sigma_{\rm fit})^2}{(d\sigma_i)^2} +
\sum_{j=1}^{N_{\rm exp}} \frac{(\eta_j - 1)^2}{(d\eta_j)^2},
\label{eq:chisq}
\end{equation}
where $\sigma_i$ and $d\sigma_i$ are the cross section and error (excluding
normalization uncertainties) for each of the $N_\sigma$ data points, $\eta_j$
is the fitted normalization factor for the $j$th dataset, and $d\eta_j$ is the
normalization uncertainty for that dataset. We fit to 470 data points
($N_\sigma$=443, $N_{exp}$=27) with 39 parameters (six parameters each for
the electric and magnetic form factors, and 27 normalization parameters).

\begin{figure}
\includegraphics[height=7.0cm,width=8.0cm,angle=0]{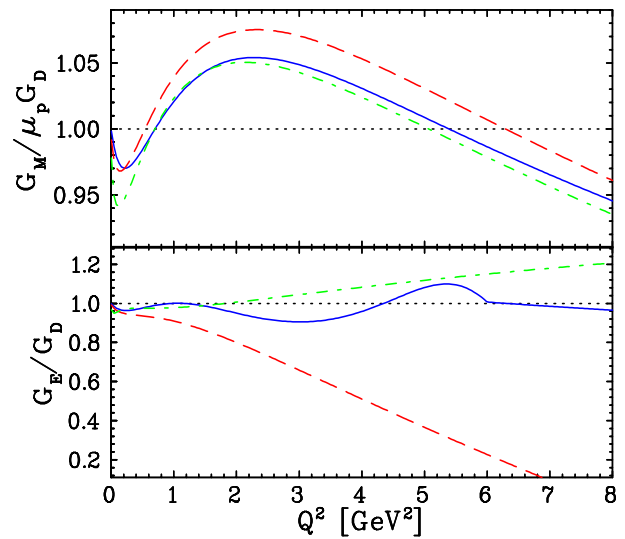}
\caption{(Color online) The ``Rosenbluth form factors'' (solid line) for
$\ge$ and $\gm$ relative to the dipole form: $G_D = [1 + Q^2/M_D^2]^{-2}$,
$M_D^2$=0.71~GeV$^2$.  The dot-dashed line is the previous fit to Rosenbluth
extracted form factors from~\cite{bosted94}, and the dashed curve is the fit to
$\gm$ from~\cite{brash02}, with the form factor ratio constrained to give
$\mugegm = 1 - 0.13(Q^2-0.04)$. \label{fig:newfit1}}
\end{figure}

\begin{table}
\caption{Fit parameters for the Rosenbluth form factors, using the
parametrization of Eq.~\ref{eq:fitform}.
%Values of $p_2$ correspond to an
%uncorrected RMS charge (magnetization) radius of 0.869 (0.863) fm.
%Values are quoted to three or four significant digits, keeping enough
%precision to make this parametrization indistinguishable from the full
%precision fit.  This can be quite important, due to the significant
%cancellation between terms, especially at large $Q^2$ values.
\label{tab:xsec_fit}}
\begin{ruledtabular}
\begin{tabular}{ccc}
Parameter & $\ge$ (Rosenbluth)	& $\gm/\mu_p$ (Rosenbluth) 	\\
\hline
$p_2$	& 3.226		& 3.19			\\
$p_4$	& 1.508		& 1.355			\\
$p_6$	& -0.3773	& 0.151			\\
$p_8$	& 0.611		& -1.14$\times 10^{-2}$	\\
$p_{10}$ & -0.1853	& 5.33$\times 10^{-4}$		\\
$p_{12}$ & 1.596$\times 10^{-2}$ & -9.00$\times 10^{-6}$	\\
\end{tabular}
\end{ruledtabular}
\end{table}

The result of the global fit to the cross section data is shown in
Fig.~\ref{fig:newfit1}.  The fit yields a total $\chi^2$ of 326.7 for 431
degrees of freedom, yielding a reduced $\chi^2$, $\chi^2_\nu =
\chi^2/N_{dof}$, of 0.758. This yields an unreasonably high confidence level,
indicating that the quoted uncertainties of the measurements are too large. 
As was observed in the previous fit~\cite{arrington03}, the majority of the
datasets, 20 out of 27, have values of $\chi^2_\nu < 1$, indicating that most
of the experiments were overly conservative in estimating their uncertainties.
Table~\ref{tab:xsec_fit} lists the parameters for the Rosenbluth form factors.
The fit includes cross sections for $Q^2$ values from 0.005 to 30~GeV$^2$, and
should be valid over this range, though the separation of $\ge$ and $\gm$ is
only well constrained by the data for $Q^2 \ltorder 6$~GeV$^2$.

The normalization factors were generally smaller than the quoted scale
uncertainties of the experiments ($\chi^2$=18.0 for 27 normalization factors).
The average normalization factor is 0.65\%, and the RMS normalization factor
is 2.7\%. The normalization factors are very close to those obtained in 
the previous global fit~\cite{arrington03}. The average normalization factor
differs by approximately 0.5\%, and the individual normalization factors
differ by less then 1\% for 18 of the 20 experiments. Because the previous
fit excluded data below $Q^2$=0.6~GeV$^2$, the agreement indicates that it is
the self-consistency, rather than the form factor constraint at $Q^2=0$, that
dominates the determination of the normalization factors.

%\begin{figure}
%\includegraphics[height=5.0cm,width=8.0cm,angle=0]{compare_xsec_660_log.ps}
%\caption{(Color online) Comparison of $\mugegm$ with single-experiment
%extractions. The red circles are the re-extracted values from
%Ref.~\cite{arrington03}, the green squares are the values from Christy,
%\etal~\cite{christy03}, and the blue triangles are from Dutta,
%\etal~\cite{dutta03}.) \label{fig:single_experiment}}
%\end{figure}
%
%Figure~\ref{fig:single_experiment} compares the results of this fit to the
%direct single-experiment Rosenbluth extractions, taken from
%Refs.~\cite{arrington03, dutta03, christy03}.  By comparing only the
%single-experiment extractions, we avoid the potentially large and correlated
%uncertainties that arise from the relative normalization of different
%datasets.  The comparison of data to fit yields $\chi^2=45.3$ for 50 data
%points ($\chi^2=17.8$ for the 20 points above $Q^2=1.5$~GeV$^2$).

We can test the self-consistency of the individual data
sets by comparing the global fit to the results of single-experiment
extractions of $\ge$ and $\gm$. By comparing only the single-experiment
extractions, we avoid the potentially large and correlated uncertainties that
arise from the relative normalization of different datasets.  Comparing
the ratio $\gegm$ from the fit to the individual experiments, taken from
Refs.~\cite{christy03, dutta03} and the reanalysis of older experiments
presented in Ref.~\cite{arrington03} yields $\chi^2=45.3$ for 50 data points
($\chi^2=17.8$ for the 20 points above $Q^2=1.5$~GeV$^2$).

\begin{figure}
\includegraphics[height=7.0cm,width=8.0cm,angle=0]{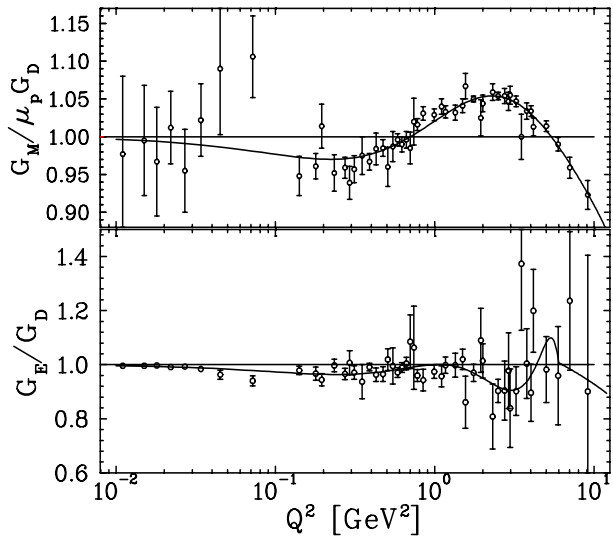}
\caption{$\gm$ (top) and $\ge$ (bottom) from direct Rosenbluth separation
utilizing normalization factors from the global fit.
\label{fig:directlt}}
\end{figure}

We can estimate the uncertainties in the form factors by performing direct
Rosenbluth separations in several $Q^2$ bins using the full dataset, with
normalization factors determined from the global fit.  For each $Q^2$ bin, the
data are scaled to the average $Q^2$ value of the data points in that bin,
using the global fit as the scaling function. $Q^2$ bins were chosen so that
there are at least three data points in the bin, the $\epsilon$ range covered
is at least 0.3, and the correction for scaling each point to the average
$Q^2$ value was $\ltorder$10\% (typically $<2$\%).  The scaling was also done
using the fits of Refs.~\cite{bosted94} and~\cite{brash02}, shown in
Fig.~\ref{fig:newfit1}.  Varying the scaling procedure changed the ratios by
$\ll 1$\%, except for the very highest (lowest) $Q^2$ points, where the
change in $\ge$ ($\gm$) was as much as 3\%, but was still much smaller than
the uncertainty in the extracted form factor.

Figure~\ref{fig:directlt} shows the fits to $\ge$ and $\gm$, along with the
direct Rosenbluth separation points, using the normalization factors from the
fit. Except for the very low $Q^2$ values, typical uncertainties on $\gm$ are
$\approx$1\%, increasing to $\sim$2\% for $Q^2$=10~GeV$^2$ (8\% for
$Q^2$=30~GeV$^2$).  At low $Q^2$, the experimental uncertainties become quite
large, but the constraint on the behavior as $Q^2 \rightarrow 0$ yields a much
smaller uncertainty on the fit. For $\ge$, the uncertainties are (1--2)\% at
low $Q^2$, but are (5--10)\% for intermediate $Q^2$ values (2--4~GeV$^2$), and
grow rapidly as $Q^2$ increases. Note that the uncertainties in $\ge$ and
$\gm$ are highly anti-correlated, due to the way the form factors are
separated from the cross section measurements.  This can be seen in the
anti-correlation of the deviation of the points from the fits in
Fig.~\ref{fig:directlt}. Thus, the uncertainty on the cross sections extracted
from this parametrization is not just the sum of the uncertainties in the
contributions from $\ge$ and $\gm$.  Up to $Q^2$$\approx$4~GeV$^2$, there is a
large body of cross section measurements with point-to-point uncertainties of
$\sim$1\%.  Because the normalization factors are determined in the fit, and
the residual uncertainty in the normalization is small, the absolute cross
sections should be known to better than 2\%.  Above $Q^2$=4~GeV$^2$, the
number of data points decreases, and the uncertainties in the cross sections
grow, reaching 10\% at $Q^2$=25~GeV$^2$.

Even with the uncertainty related to the discrepancy between Rosenbluth
and polarization transfer, this fit yields a precise parametrization of
the elastic cross section in the one-photon exchange formalism. While 
these may not be the underlying form factors of the proton (e.g., if there are
missing radiative correction terms), this is still the appropriate
parametrization to use as input to a calculation or analysis that requires
the elastic cross section.  Using the form factors derived from the
polarization transfer technique will not yield the correct cross section,
even in a combined analysis of Rosenbluth and polarization transfer such
as performed in Refs.~\cite{brash02, arrington03}.  More importantly, an
{\it inconsistent} combination of cross section and polarization transfer
results can magnify the error.  Combining a parametrization of $\gm$ from
a Rosenbluth analysis with the form factor ratios measured in polarization
transfer decreases $\ge$, and thus decreases the total cross section, relative
to the best fit to the cross section data, without allowing a corresponding
increase in $\gm$.  This leads to form factors which give cross sections that
are (4--10)\% below the measured cross sections at large $\epsilon$ over a
large $Q^2$ range ($0.1<Q^2<15$~GeV$^2$).

\section{COMBINED POLARIZATION AND ROSENBLUTH ANALYSIS}\label{sec:combined}

While the Rosenbluth form factors yield the best parametrization for the cross
section in the usual one-photon exchange picture, the ratio does not agree
with the ratio extracted from the polarization transfer technique.  For the
larger $Q^2$ values, the polarization transfer technique is less
sensitive to knowledge of the kinematics, radiative correction, and other
systematic uncertainties that are important in the Rosenbluth separation.

If this discrepancy is related to a problem in the cross section data, then
the polarization transfer will yield the true ratio of the form factors,
but has to be combined with cross section data to obtain both $\ge$ and
$\gm$.  We present in this section a combined analysis of the polarization
transfer and cross section data, which will yield the ``Polarization form
factors''.

In order to obtain a consistent extraction of the form factors, we must make
an assumption about the nature of the discrepancy.  We assume that the
difference comes from a common systematic error in the cross section
measurements. Analyses of this discrepancy~\cite{arrington03, guichon03}
indicate that there must be an $\epsilon$-dependent correction of
(5--7)\%, roughly linear in $\epsilon$, for $1 < Q^2 <6$ GeV$^2$.

In the combined analysis, we apply a linear, $\epsilon$-dependent correction
of 6\%, to all datasets.  This is the `minimal' assumption necessary to
make the two techniques consistent, to the extent that a correction that was
not linear in $\epsilon$, or which modified only some of the datasets, would
have to be larger. A correction that is nearly linear in $\epsilon$ and
fairly $Q^2$-independent is consistent with the form for the two-photon
exchange term in the analysis of Ref.~\cite{blunden03}, although the
size of the correction in Ref.~\cite{blunden03} is only $\sim$2\%, less than
half the size necessary to explain the discrepancy.

We repeat the fit from Sec.~\ref{sec:xsec}, but with cross sections
modified by the linear $\epsilon$ dependence, and with the polarization
transfer data included in the fit, as described in~\cite{arrington03}.  The
correction to the cross section could either lower the cross section at large
$\epsilon$ values, or increase it at small $\epsilon$ values:
\begin{eqnarray}
\label{eq:corr1}
\sigma_{c1} = \sigma_0 (1 - 0.06 \epsilon ) \\
\label{eq:corr2}
\sigma_{c2} = \sigma_0 (1 - 0.06 (\epsilon-1)) = 0.94 \sigma_{c1}
\end{eqnarray}
The first correction is consistent with the form from Ref.~\cite{guichon03},
while the second is consistent with the behavior of~\cite{blunden03}.
The second form was chosen for the main fit because the correction is small at
large $\epsilon$ (small $\theta_e$), where comparisons of positron to electron
scattering from SLAC~\cite{mar68} set fairly tight limits on the size of
two-photon exchange.

\begin{figure}
\includegraphics[height=7.0cm,width=8.0cm,angle=0]{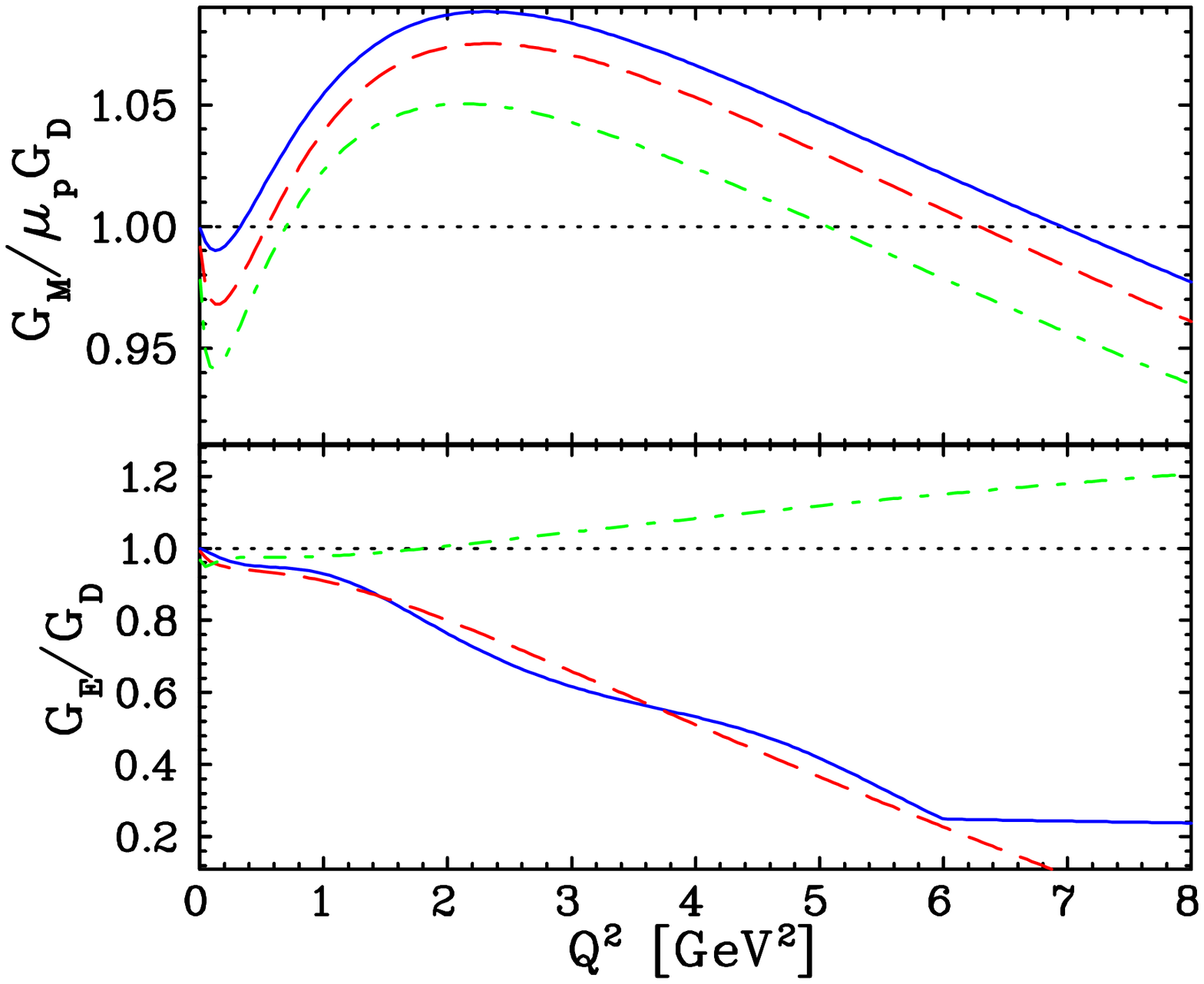}
\caption{(Color online) The ``Polarization form factors'' (solid line) for
$\ge$ and $\gm$, relative to the dipole form.  The dot-dashed line is the
previous fit to Rosenbluth extracted form factors from~\cite{bosted94}, and
dashed curve is the fit to $\gm$ from~\cite{brash02}, with the form factor
ratio constrained to give $\mugegm = 1 - 0.13(Q^2-0.04)$. \label{fig:newfit2}}
\end{figure}

\begin{table}
\caption{Fit parameters for the Polarization form factors, using the
parametrization of Eq.~\ref{eq:fitform}.
\label{tab:combined_fit}}
\begin{ruledtabular}
\begin{tabular}{ccc}
Parameter & $\ge$ (Polarization)	& $\gm/\mu_p$ (Polarization) 	\\
\hline
$p_2$	& 2.94		& 3.00		\\
$p_4$	& 3.04		& 1.39		\\
$p_6$	& -2.255	& 0.122		\\
$p_8$	& 2.002		& -8.34$\times 10^{-3}$	\\
$p_{10}$& -0.5338 	&  4.25$\times 10^{-4}$	\\
$p_{12}$& 4.875$\times 10^{-2}$ & -7.79$\times 10^{-6}$	\\
\end{tabular}
\end{ruledtabular}
\end{table}

The Polarization form factors, from the combined fit to the cross section
and the 26 polarization transfer data points from Refs.~\cite{jones00,
gayou01, gayou02} is shown in Fig.~\ref{fig:newfit2}.  The fit yields a total
$\chi^2$ of 391.6 for 457 degrees of freedom, $\chi^2_\nu=0.857$, including
the additional $\chi^2$ contribution for the polarization transfer data (Eq.
(8) of Ref.~\cite{arrington03}). Table~\ref{tab:combined_fit} gives the fit
parameters for the Polarization form factors.

The fit was also performed with the correction of Eq.~\ref{eq:corr1}.
This leads to an overall rescaling of all of the cross sections by 6\%,
relative to the correction of Eq.~\ref{eq:corr2}. However, this does not
yield a simple rescaling of the form factors, because each dataset has
a normalization factor that is determined in the fit, and because the
form factors are constrained to reproduce the charge and magnetic moment at
$Q^2=0$.  While a two-photon correction of this size for large $\epsilon$
values would appear to be ruled out by the SLAC positron-proton
measurements~\cite{mar68}, an $\epsilon$-dependent systematic other than
two-photon exchange could also resolve the discrepancy. However, this fit
yields a much worse $\chi^2$ value: 575.1 for 457 degrees of freedom, and
so we choose to apply Eq.~\ref{eq:corr2} for our combined fit.

Note that the result of the combined fit (Table~\ref{tab:combined_fit})  will
\textit{not} reproduce the measured elastic cross section in the one-photon
exchange formalism; it will reproduce the \textit{modified} cross sections of
Eq.~\ref{eq:corr2}. Therefore, the Polarization form factors should not be
used to model elastic electron-proton cross section measurements.  However, if
the `minimal' assumption that a correction consistent with the form of
Eq.~\ref{eq:corr2} explains the discrepancy, this should yield a consistent
extraction of the underlying form factors of the proton.

\section{CONCLUSIONS}

Form factors extracted using the Rosenbluth technique provide a parametrization
of the deviation of the elastic electron-proton cross section from the
point-scattering cross section. If the cross section has additional
corrections, such as two-photon exchange terms, that are not being taken into
account, then the Rosenbluth extraction does not yield the true proton form
factors that relate to the structure of the proton. In this case, $\ge$ must
be extracted from the polarization transfer measurements, which yield $\gegm$,
and the cross section data must be utilized to determine $\gm$. While we
cannot know how to properly combine the polarization transfer and cross
section data until we understand the cause of the discrepancy, the
uncertainties in $\gm$ that arise from this problem are much smaller than
those in $\ge$. The same holds true if there is some other correction or
combination of corrections to the cross section other than the two-photon
exchange (e.g. Coulomb corrections~\cite{higinbotham03}). It is of course
possible that the discrepancy is due to a problem with the polarization
transfer data or technique rather than the cross section data. If so, then the
Rosenbluth form factors represent both the correct cross section and the
correct nucleon structure.  However, there do not appear to be any obvious
candidates for problems in the technique, and the experiment should be less
prone to systematic uncertainties than the Rosenbluth extractions.

We have presented two extractions of the proton electromagnetic form factors.
The Rosenbluth form factors come from a global Rosenbluth extraction of
the form factors from electron-proton elastic scattering measurements. The
Polarization form factors come from a combined fit to the cross section
and polarization transfer data, under the assumption that the discrepancy
between the techniques is caused by a linear, $\epsilon$-dependent correction
to the cross sections.  The Rosenbluth form factors give a global
parametrization of the elastic electron-proton scattering cross section in the
one-photon exchange approximation.  Even if there is a correction to the cross
sections, neglected in the one-photon exchange formalism, this parametrization
will yield the correct cross sections in the one-photon approach.  Under the
above assumption of an unknown correction to the cross sections, the
Polarization form factors yield the underlying form factors, but will not
reproduce cross sections, and will therefore yield incorrect results if used
as input for an analysis that requires the elastic cross section, as was
observed in an analysis of quasielastic scattering from nuclei~\cite{dutta03}.

Additional data will help shed light on the origin of the discrepancy.  An
improved ``Super-Rosenbluth'' separation measurement~\cite{e01001} completed
at Jefferson Lab in 2002 will yield a precise extraction of $\gegm$, and
will determine if the discrepancy can be explained by experimental problems in
the Rosenbluth extractions.  A new polarization transfer
experiment~\cite{e01109}, approved to run at Jefferson Lab, will provide an
independent confirmation of the existing polarization transfer results, as
well as extending the measurements to higher $Q^2$ values.  If sufficiently
improved calculations or direct measurements of the two-photon exchange
corrections become available, we should be able to determine if they are
responsible for the discrepancy, and if so, remove the current uncertainty in
combining cross section and polarization transfer measurements.

If the discrepancy is explained by two-photon corrections or some other
effect on the cross sections, and we have reliable calculations for these
effects, then the cross section data can be combined with the polarization
transfer data to extract the form factors without ambiguity.  These form
factors will represent the underlying structure of the proton and provide a
useful parametrization of the elastic electron-proton cross section, as long
as the effect is properly accounted for.  Until the discrepancy is well
understood, however, both sets of form factors are necessary, and it is
important to use form factors that are (1) extracted consistently from the
cross section and/or polarization transfer data, and (2) appropriate for the
problem being addressed.

\begin{acknowledgments}

This work was supported by the U. S. Department of Energy, Nuclear Physics
Division, under contract W-31-109-ENG-38.

\end{acknowledgments}

\bibliography{protonff}

\end{document}